\documentclass[twocolumn,aps,showpacs,superscriptaddress,tightenlines,a4paper]{revtex4}
\pdfoutput=1
\usepackage{amssymb}
\usepackage{amsmath}
\usepackage{bbold}
\usepackage{mathrsfs}
\usepackage{paralist}
\usepackage{graphicx}
\usepackage{xcolor}

\renewcommand{\vec}[1]{\mathbf{#1}}

\begin{document}
%
%
	\title{The paradox of the Casimir force in inhomogeneous transformation media}
%
%
	\author{William M. R. Simpson}
	\affiliation{School of Physics and Astronomy, University of St Andrews, North Haugh, St Andrews, KY16 9SS, UK}
	\affiliation{Department of Physics of Complex Systems, Weizmann Institute of Science, Rehovot 76100, Israel}

%
%
	\begin{abstract}
	It has recently been argued that Casimir-Lifshitz forces depend in detail on the microphysics of a system; calculations of the Casimir force in inhomogeneous media yield results that are cutoff-dependent. This result has been shown to hold generally \cite{simpson2013,horsleysimpson2013}. But suppose we introduce an inhomogeneous metamaterial into a cavity that effectively implements a simple distortion of the coordinate system. Considered in its `virtual space', the optical properties of such a material are homogeneous and consequently free from the cutoff-dependency associated with inhomogeneous media. This conclusion should be reconciled with recent advances in our understanding of Casimir-Lifshitz forces. We consider an example of such a system here and demonstrate that, whilst the size of the Casimir force is modified by the inhomogeneous medium, the force is cutoff-independent and can be stated exactly. The apparent paradox dissolves when we recognise that an idealised metamaterial that could implement a virtual geometry for all frequencies would be devoid of internal scattering, and would not give rise to a cutoff-dependency in the Casimir force for that reason.
	\end{abstract}

	%
	%
	\maketitle
	\bibliographystyle{unsrt}
%
%

\section{Introduction}

	\par
	Casimir-Lifshitz forces arise from the ground-state properties of the electromagnetic field, and have been computed for a variety of systems and geometries (\cite{lifshitz1955,DzyLifPit1961,volume9,philbin2011}). Although most analytic calculations of the Casimir force involve idealisations, it is possible to predict the forces in more realistic models, where effects such as dispersion and dissipation are properly taken into account. Lifshitz theory offers such an apparatus \cite{DzyLifPit1961}. However, it still remains beyond the scope of present theory to predict both the nature and size of Casimir forces in many simple systems. For example, consider the case of a cavity with perfect mirrors -- Casimir's original model. It is possible to calculate a Casimir force for an empty cavity \cite{casimir1948}, or a cavity filled with perfectly homogeneous fluid, such as purified water \cite{lifshitz1955,capasso2009}. A spoonful of sugar dissolved in the water, however, is enough to frustrate the calculation; Lifshitz theory (as well as less sophisticated methods) predicts an infinite force on the mirrors with even the gentlest perturbation in the optical properties of the medium, if the perturbation is described as a continuous function of position (such as a sugar solution under gravity). We have found that Casimir forces are, in general, impossible to predict in inhomogeneous media without incorporating further detail about the microphysics of the system \cite{simpson2013,horsleysimpson2013}. The procedure for doing so has yet to be explicated \footnote{In previous work \cite{simpson2013,horsleysimpson2013}, it has been speculated that the solution to this problem may involve taking spatial dispersion into account}.

	\par
	However, as an idealised thought-experiment, we can imagine introducing an inhomogeneous metamaterial into a cavity whose effect on light could be interpreted as a simple distortion of the laboratory coordinate system. Such media are common to the field of transformation optics, and have been put to use in various applications (such as simple cloaking devices, for example) \cite{leonhardt_geometry_2010,pendry__2006}.  Considered in the `virtual space'~\footnote{Maxwell's equations in a transformed coordinate system (transformed from a Cartesian grid) in empty space take the same form as Maxwell's equations in a Cartesian system with a medium present. These different pictures are referred to as the `virtual' and the `physical' space respectively~\cite{leonhardt_geometry_2010}. In swapping between them, the physics should remain the same.} of the geometry that such a device effectively implements, the coordinate transformed values of the permittivity and permeability tensors within the device remain homogeneous. In the coordinate system of physical space, however, the optical properties of the device vary continuously. There appears to be a contradiction, then, between the results we obtain from transformation optics, when the problem is calculated in its virtual space, and the force that Lifshitz theory predicts in the physical space of the laboratory, which is presumably cutoff-dependent \cite{simpson2013,horsleysimpson2013}. This is paradoxical: simply changing our preferred coordinate system cannot change a force of nature that exists between the plates.

	\par
	In fact there is no contradiction here, when the locus of the problem is specified more precisely: The Casimir force depends on the microphysical details of scattering within a system, and inhomogeneous media typically give rise to internal scattering. In such systems, the Casimir force is cutoff--dependent. However, in the peculiar case of optical devices that implement virtual geometries for light, no additional scattering is introduced into the system. Such a device can modify the size of the Casimir force, as we demonstrate, but these effects can be computed without reference to the microphysical details of the system or making recourse to any additional physics.

%
%
\section{The Casimir Force in an empty cavity}
	\par
	Before considering the details of the system we will examine, it is instructive to return to the simple case of the Casimir force in an empty cavity composed of two perfect mirrors~\cite{casimir1948}.  In Casimir's original calculation, the force was deduced from a mode summation of the ground-state energy of the electromagnetic field. The same result can be recovered using the more general formalism of Lifshitz theory. The formalism, in this case, is written in terms of the electromagnetic Green function, which describes the field produced by sources of current within the system. The ground state of the coupled system of electromagnetic field and dielectric is one with non–zero current density within the media~\cite{philbin2010,philbin2011}, consistent with the fluctuation dissipation theorem~\cite{volume5}. Casimir–-Lifshitz forces can be understood as arising from the interaction of such currents. A stress tensor is written in terms of this Green function, from which the force can be derived.
	\par
	However, for a region or cavity of width \(d\) where \(\epsilon\) and \(\mu\) are homogeneous, the value of the regularised stress tensor at a point \(x\) can be written in terms of the reflection coefficients associated with sending \(q\)--polarized (\(q=s,p\)) plane waves to the right (\(r^{q}_{R}\)) and to the left (\(r^{q}_{L}\)) of this point~\cite{lifshitz1955,genet2003,leonhardt2010},
	\begin{equation}
		\sigma_{xx}(x)=2\hbar c\sum_{q=s,p}\int_{0}^{\infty}\frac{d\kappa}{2\pi}\int_{\mathbb{R}^{2}} \frac{d^{2}\boldsymbol{k}_{\parallel}}{(2\pi)^{2}}w\frac{r^{q}_{L}r^{q}_{R}e^{-2d w}}{1-r^{q}_{L}r^{q}_{R}e^{-2d w}}\label{casimir-stress-tensor}
	\end{equation}
	where \(w=(n^{2}\kappa^{2}+k_{\parallel}^{2})^{1/2}\), \(k_{\parallel}=|\boldsymbol{k}_{\parallel}|\), and \(n\) is the value of the refractive index in the homogeneous region surrounding \(x\).  The reflection coefficients are functions of the imaginary frequency, \(\omega=ic\kappa\), the (real) in--plane wave--vector \(\boldsymbol{k}_{\parallel}\), and the material parameters of the media to the right and to the left of the homogeneous region. In an isotropic system, they take the form
\begin{align}
r^{s}_{L} &=& & \frac{\mu_{L}w-\mu w_{L}}{\mu_{L}w-\mu w_{L}},&\quad
r^{s}_{R} &=& &\frac{\mu_{R}w-\mu w_{R}}{\mu_{R}w+\mu w_{R}},& \nonumber
\\
r^{p}_{L} &=& -&\frac{\epsilon_{L}w-\epsilon w_{L}}{\epsilon_{L}w-\epsilon w_{L}},&\quad
r^{p}_{R} &=& -&\frac{\epsilon_{R}w-\epsilon w_{R}}{\epsilon_{R}w+\epsilon w_{R}},& \label{eq:Fresnel}
\end{align}
where the $L$ subscript indexes properties to the left of the region or cavity, and the $R$ subscript indexes properties to the right. The advantage of writing the stress tensor in this form is that the regularization procedure of Lifshitz theory is automatically implemented~\cite{leonhardt2010}; the contributions to the stress arise entirely from the scattering properties of the system.
	\par
	In Casimir's case, where the cavity is composed of two perfect mirrors with vacuum between, we take the limit $\epsilon_{L}, \epsilon_{R} \rightarrow \infty $ with $ \mu_{L} = \mu_{R} = 1$. Since there is only vacuum between the plates, $ \epsilon = \mu = 1$. Thus
\begin{equation}
r_{L} = r_{R} = -1. \label{eq:plate-coeffs}
\end{equation}
Following the procedure outlined in~\cite{leonhardt2010}, the integral above can be reexpressed as
\begin{align}
\sigma_{xx} &=& &\frac{\hbar c}{2\pi^{3}}\int_{0}^{\infty}\int_{0}^{2\pi}\int_{0}^{\pi/2}\frac{w^{3}\text{sin}\theta\:\text{d}\theta\text{d}\phi\text{d}w}{\text{e}^{2d w}-1}& \nonumber \\
&=& &\frac{\hbar c}{\pi^{2}}\int_{0}^{\infty}\frac{w^{3}}{\text{e}^{2d w}-1},&
\end{align}
from which we derive the usual expression for the Casimir pressure \cite{casimir1948}:
\begin{equation}
P=-\frac{\hbar c\pi^{2}}{240 d^{4}}. \label{eq:Casimir-force}
\end{equation}

%
%
\section{The Casimir force in a 'compressed' cavity}

\subsection{The C-Slice}

	\par
	We consider now a simple modification of the system described above, where a wafer characterised by   anisotropic, inhomogeneous electric permittivity and magnetic permeability tensors
\begin{equation}
\epsilon(z)=\mu(z)=\left(\begin{array}{ccc}
m(z,\omega)^{-1} & 0 & 0\\
0 & m(z,\omega)^{-1} & 0\\
0 & 0 & m(z,\omega)
\end{array}\right) \label{eq:specs}
\end{equation}
is inserted between the perfect mirrors of the cavity. The cavity mirrors are positioned at $z=0, d$, and the wafer occupies an interval $z \in[a,b] \subset[0,d]$.  The wafer, which we will refer to as a \emph{C-}-slice \footnote{The C here denotes a \emph{compression} of the vacuum. The compressive properties of a homogeneous, non-dispersive form of this metamaterial have been discussed pedagogically in \cite{kundtz2011}.}, is impedance-matched and is intended to implement a coordinate transformation in which an interval of space is compressed along the $z-$axis by a factor
\begin{equation}
C_{S}=\frac{1}{\Delta}\int_{a}^{b}m(z,\omega)\,\text{d}x,\quad \Delta = b-a. \label{eq:compression}
\end{equation}
It is not difficult to see why the $C-$slice should have these properties \cite{kundtz2011}: Beginning first with a homogeneous, non-dispersive $C-$slice, where the compression can be characterised by a simple factor $m(z,\omega)=const=m$, we consider an electromagnetic wave with wave number $k$ travelling parallel to the intended axis of compression. The wave must pick up the same phase, as it passes through the length of the device, as it would have acquired had it passed through an uncompressed region. Hence
\begin{equation}
k \Delta = k m \Delta \sqrt{\epsilon_{\perp} \mu_{\perp}}. \label{eq:phasematch}
\end{equation}
Here, $\epsilon_{\perp}$ and $\mu_{\perp}$ are the permittivity and permeability components normal to the axis of compression; $\epsilon_{\parallel}$ and $\mu_{\parallel}$ are the components that are parallel to it. From (\ref{eq:phasematch}) we deduce that
\begin{equation}
\epsilon_{\perp} \mu_{\perp} = m^{-2} \label{eq:phasematch2}.
\end{equation}
In order to be equivalent to a distortion of the coordinate system, we require impedance-matched media, $\epsilon = \mu$, hence
\begin{equation}
\epsilon_{\perp} = \mu_{\perp} = m^{-1}. \label{eq:impmatch}
\end{equation}
Consider now a wave travelling along the $x-$axis, perpendicular to the axis of compression, and therefore free of any compression, polarised so that the magnetic field lies along the $z-$axis, and the electric field along the $y-$axis. In this case
\begin{equation}
k = k \sqrt{\epsilon_{\perp} \mu_{\parallel}} \label{eq:phasematch2},
\end{equation}
from which we deduce that $\epsilon_{\parallel} = m$, and hence $\epsilon_{\parallel} = m$. The generalisation to the inhomogeneous case is trivial: for each infinitesimal region $\delta z$, the space is now compressed to $m(z)^{-1} \delta z$, and the overall compression factor is given by (\ref{eq:compression}). Finally, by incorporating a frequency dependence, $m(z,\omega)$, we allow different frequencies to experience different levels of compression, and therefore allow for some degree of dispersion.
	\par
	However, this remains an idealised model. The Casimir Effect is a broadband phenomenon, and it is difficult to see how the necessary condition of impedance-matching could in practice be secured for a sufficiently large portion of the electromagnetic spectrum~\footnote{For a more positive appraisal of the utility of metamaterials in Casimir Physics experiments, see~\cite{leonhardt2007}}. Nevertheless, the practical difficulties of implementing this device should not disqualify its use in a {\it Gedankenexperiment}. We will briefly discuss some possible applications for the $C-$slice later in this paper.

\subsection{The Casimir force in virtual space}\label{subsec-virtualcalc}

	\par
In the `virtual space' of the $C-$slice, the transformed values of the permittivity and permeability tensors are equal to unity, ie. the properties of vacuum. The addition of the wafer simply modifies the effective length of the cavity, seen by a given frequency $\omega$, from a distance $d$ to
\[
d' = d + \Delta(C_{S}^{-1} - 1).
\]
For this reason, we have grounds for expecting the force between the plates, with or without the $C-$slice, to remain cutoff-independent. If we ignore dispersion by setting $m(z,\omega)=m(z)$, it is clear that the modified Casimir pressure can be stated exactly by simply substituting the distance paramater $d$, in the original expression for the Casimir force (\ref{eq:Casimir-force}), with the effective length of the cavity, $d'$:
\begin{equation}
P=-\frac{\hbar c\pi^{2}}{240 d'^{4}} = -\frac{\hbar c\pi^{2}}{240 \left( d + \Delta(C_{S}^{-1} - 1) \right)^{4}}.
\label{eq:newCasimir}
\end{equation}

But how does this compare with a treatment of the problem by Lifshitz theory within its `physical space'?

\subsection{The Casimir force in physical space}

\subsubsection{Lifshitz theory in anisotropic media}

	\par
	The vacuum stress of the cavity, according to Lifshitz theory, is determined by the scattering properties of its constituents, and can be codified in the form of reflection coefficients. In the case of anisotropic media, the form of the stress must be modified~\cite{rosa2008}, replacing (\ref{casimir-stress-tensor}) with
\begin{equation}
		\sigma_{xx}(x)=2\hbar c\int_{0}^{\infty}\frac{d\kappa}{2\pi}\int_{\mathbb{R}^{2}} \frac{d^{2}\boldsymbol{k}_{\parallel}}{(2\pi)^{2}}w\text{Tr}\frac{\vec{R}_{L} \vec{R}_{R}e^{-2d w}}{1-\vec{R}_{L} \vec{R}_{R} e^{-2d w}}\label{casimir-stress-tensor2},
\end{equation}
where
\begin{equation}
\vec{R}_{L} = \left(\begin{array}{ccc}
r^{ss}_{L} & r^{sp}_{L} \\
r^{ps}_{L} & r^{pp}_{L}
\end{array}\right), \quad
\vec{R}_{R} = \left(\begin{array}{ccc}
r^{ss}_{R} & r^{sp}_{R} \\
r^{ps}_{R} & r^{pp}_{R}
\end{array}\right)\label{eq:reflecmatrices},
\end{equation}
and $r^{pq}_{L,R}$ is the ratio of a field with $p-$polarization divided by an incoming field with $q-$polarization, for reflection from the left (L) or right (R). The indices $s$ and $p$ correspond respectively to perpendicular and parallel polarizations with respect to the plane of incidence. In the case of isotropic media, the off-diagonal elements of both matrices vanish and the diagonal elements are given by the familiar Fresnel expressions (\ref{eq:Fresnel}). The expression above then reduces to the usual Lifshitz formula (\ref{casimir-stress-tensor}).

\subsubsection{Transfer matrices}

To determine how a $C-$slice modifies the Casimir force, we must determine the reflection coefficients of the device, for all angles of incidence. The $C-$slice we are considering, in this case, is an inhomogeneous material; its optical properties vary continuously along the $z-$axis. A set of reflection coefficients may be obtained, however, by approximating the cavity as a series of $N+1$ homogeneous slices of width $\delta z$, where each slice has electric permittivity $\epsilon_{j}$ and magnetic permeability $\mu_{j}$, and $j\in [0,N+1]$ indexes the slicing. Between the plates, the electric and magnetic response is characterised by
\begin{equation}
\epsilon_{j}=\mu_{j}=\left(\begin{array}{ccc}
m_{j}^{-1} & 0 & 0\\
0 & m_{j}^{-1} & 0\\
0 & 0 & m_{j}
\end{array}\right)\label{eq:C-Slice-slice}\quad \text{for}\,j \in [2,N].
\end{equation}
For an empty slice, $m_{j}=1$. Once the reflection coefficients have been determined, the stress may be calculated. The transfer matrix technique can be used for such an analysis of the field~\cite{born1999,genet2003,shelykh2004,artoni2005,hao2008}, the field in strip \(j+1\) being related to the field in \(j\) by
	\begin{equation}
		\boldsymbol{E}(j+1)=\boldsymbol{t}(j+1)\boldsymbol{\cdot}\boldsymbol{E}(j)\label{transfer-matrix},
	\end{equation}
	where $\boldsymbol{t}(j+1)$ is the $4 \times 4$ transfer matrix relating the field on the far right of slice \(j\) to that on the far right of slice \(j+1\), and $\boldsymbol{E}(j)$ is a set of expansion coefficients for the field in slice $j$. The transfer matrix can be decomposed into two components,
\begin{equation}
 \vec{t}(j+1) = \vec{M}(j+1)\, \vec{\Phi}(j+1),
\end{equation}
where $\vec{M}(j+1)$ implements the boundary conditions at the interface between slices $j$ and $j+1$, derived under the condition that the tangential components of the field should be continuous, and $\vec{\Phi}(j+1)$ is a diagonal matrix consisting of phase propagation terms that evolves the field between the two boundaries. To determine $\vec{M}(j+1)$, we compute 
\[
 \vec{M}(j+1)=\vec{D}(j+1)^{-1}\vec{D}(j),
\]
for a system of two half-spaces, where $\vec{D}(j)$ characterises the
state of the electromagnetic field on the left, and $\vec{D}(j+1)$ the medium on the right.  The matrix $\vec{D}(j)$ is defined as 
\[
\vec{D}(j)=\left(\begin{array}{cccc}
\vec{e}^{(1)}_{j}\cdot\vec{\hat{y}} & \vec{e}^{(2)}_{j}\cdot\vec{\hat{y}} & \vec{e}^{(3)}_{j}\cdot\vec{\hat{y}} & \vec{e}^{(4)}_{j}\cdot\vec{\hat{y}}\\
\vec{h}^{(1)}_{j}\cdot\vec{\hat{x}} & \vec{h}^{(2)}_{j}\cdot\vec{\hat{x}} & \vec{h}^{(3)}_{j}\cdot\vec{\hat{x}} & \vec{h}^{(4)}_{j}\cdot\vec{\hat{x}}\\
\vec{h}^{(1)}_{j}\cdot\vec{\hat{y}} & \vec{h}^{(2)}_{j}\cdot\vec{\hat{y}} & \vec{h}^{(3)}_{j}\cdot\vec{\hat{y}} & \vec{h}^{(4)}_{j}\cdot\vec{\hat{y}}\\
\vec{e}^{(1)}_{j}\cdot\vec{\hat{x}} & \vec{e}^{(2)}_{j}\cdot\vec{\hat{x}} & \vec{e}^{(3)}_{j}\cdot\vec{\hat{x}} & \vec{e}^{(4)}_{j}\cdot\vec{\hat{x}}
\end{array}\right),
\]
where $\vec{e}_{i}(j)$ and $\vec{h}_{i}(j)$ are the $i$th eigenmodes of the electric and magnetic fields respectively in a medium with the optical properties of slice $j$.  There are four eigenmodes of the field to a slice, corresponding to two independent modes propagating forwards and backwards; the general solution for the field is a linear combination of all four of them. To determine $\vec{D}(j)$, we must therefore determine each of the electric and magnetic eigenmodes of the field for slice $j$. To calculate the value of (\ref{casimir-stress-tensor2}) at a fixed point in the medium, \(x_{l}\), that is within the \(l\)th slice, we require expressions for both $\vec{R}_{R}$, and $\vec{R}_{L}$.  These can be calculated in terms of the transfer matrices
	\begin{align}
		\boldsymbol{T}_{L}&=\prod_{j=1}^{l}\boldsymbol{t}(j)\nonumber,\\
		\boldsymbol{T}_{R}&=\prod_{j=l+1}^{N+1}\boldsymbol{t}(j),
		\label{total-transfer-matrices}
	\end{align}
associated respectively with propagation through the medium to the right and to the left of \(x_{l}\). These transfer matrices determine the relative magnitudes of the field components in each slice, and therefore fix the values of the reflection coefficients (which are simply ratios of these terms).

{\em Prima facie}, the continuum case is recovered in the limit as $\delta z \rightarrow 0$ and $N \rightarrow \infty$. But here's the rub: it is precisely in this limit that the stress has been found to diverge~\cite{simpson2013,horsleysimpson2013}, and this is the crux of the paradox we seek to address! For the moment, we will blithely ignore this objection and proceed by determing the relevant transfer matrices.

\subsubsection{The boundary conditions}

The system we are considering consists of a series of homogeneous slices. The interface between two slices occurs in the $xy$--plane. Without loss of generality, we can rotate the $x$ and $y$
axes of our coordinate system so that the plane of incidence is the
$xz$ plane. Consequently the wave vectors have zero $y$ components:
\[
\vec{k}_{1}=(k_{1x},0,k_{1z}), \quad\vec{k}_{2}=(k_{2x},0,k_{2z}),
\]
where $\vec{k}_{1}$ is the wave vector of the incident light, and
$\vec{k}_{2}$ is the wave vector of the transmitted light. The frequencies
of the reflected and transmitted waves must be the same as that of
the incident wave, because the above conditions hold at the boundary
at all times -- this is only possible if the waves on either side
are oscillating at the same frequency. Additionally, the conditions
hold at all points on the boundary plane $z=0$, so the changing phases
of the waves on either side must agree as one moves along the boundary,
ie.
\begin{equation}
\vec{k}_{1}\cdot\left.\vec{r}\right|_{z=0}=\vec{k}_{2}\cdot\left.\vec{r}\right|_{z=0}.\label{eq:phase-equality}
\end{equation}
As $\vec{r}=(x,y,z),$ and the above equations hold for all values
of $x$ and $y$.
\begin{equation}
k_{1x}=k_{2x}.\label{eq:k1x=00003Dkx2}
\end{equation}
The first component of the transmitted wave vector is therefore determined
by the first component of the incident wave vector: 
\[
\vec{k}_{2}=(k_{1x},0,k_{2z}).
\]
The $k_z$ component of the wave vector, in subsequent slices, is determined by the wave equation.

\subsubsection{The wave equation}

The electric field in a homogeneous slice of the system is of the form
\begin{equation}
\vec{E}=(E_{x}\vec{\hat{x}}+E_{y}\vec{\hat{y}}+E_{z}\vec{\hat{z}})e^{i(\omega t-\vec{k}\cdot\vec{r})}.\label{eq:Eform-1}
\end{equation}
First, we derive a wave equation of general applicability. Applying Maxwell's equations
\[
\nabla\times\vec{E}=-\frac{\partial\vec{B}}{\partial t},\quad\nabla\times\vec{H}=\frac{\partial\vec{D}}{\partial t},
\]
to equation (\ref{eq:Eform-1}), we arrive at the following form of the wave equation:
\[
\left(\mbox{\boldmath$\epsilon$} \right)^{-1}\left\{ \vec{k}\times\left[\left(\vec{\mbox{\boldmath$\mu$}}\right)^{-1}\cdot\left(\vec{k}\times\vec{E}\right)\right]\right\} +\omega^{2}\vec{E}=0.
\]
This can be rewritten in matrix form. For an anisotropic metamaterial, characterised by
\[
\epsilon=\left(\begin{array}{ccc}
\epsilon_{x} & 0 & 0\\
0 & \epsilon_{y} & 0\\
0 & 0 & \epsilon_{z}
\end{array}\right),\quad\mu=\left(\begin{array}{ccc}
\mu_{x} & 0 & 0\\
0 & \mu_{y} & 0\\
0 & 0 & \mu_{z}
\end{array}\right),
\]
we obtain (\ref{eq:wavematrix1}). For a portion of an $C-$slice, characterised by (\ref{eq:C-Slice-slice}), we obtain (\ref{eq:wavematrix2}). Note that, by setting $m_{j}=1$, for all subsequent results involving $C-$slice material, we recover the properties of the vacuum.
\begin{widetext}
\begin{align}
\left(\begin{array}{ccc}
\omega^{2}\epsilon_{x}-k_{y}^{2}\mu_{z}^{-1}-k_{z}^{2}\mu_{y}^{-1} & k_{y}k_{x}\mu_{z}^{-1} & k_{z}k_{x}\mu_{y}^{-1}\\
k_{x}k_{y}\mu_{z}^{-1} & \omega^{2}\epsilon_{y}-k_{x}^{2}\mu_{z}^{-1}-k_{z}^{2}\mu_{x}^{-1} & k_{z}k_{y}\mu_{x}^{-1}\\
k_{x}k_{z}\mu_{y}^{-1} & k_{y}k_{z}\mu_{x}^{-1} & \omega^{2}\epsilon_{z}-k_{x}^{2}\mu_{y}^{-1}-k_{y}^{2}\mu_{x}^{-1}
\end{array}\right)\left(\begin{array}{c}
E_{x}\\
E_{y}\\
E_{z}
\end{array}\right)=\vec{0}. \label{eq:wavematrix1} \\
\left(\begin{array}{ccc}
-\frac{k_{y}^{2}}{m}-k_{z}^{2}m+\frac{\omega^{2}}{m} & \frac{k_{x}k_{y}}{m} & k_{x}k_{z}m\\
\frac{k_{x}k_{y}}{m} & -\frac{k_{x}^{2}}{m}-k_{z}^{2}m+\frac{\omega^{2}}{m} & k_{y}k_{z}m\\
k_{x}k_{z}m & k_{y}k_{z}m & -k_{x}^{2}m-k_{y}^{2}m+m\omega^{2}
\end{array}\right)\left(\begin{array}{c}
E_{x}\\
E_{y}\\
E_{z}
\end{array}\right)=\vec{0} \label{eq:wavematrix2}
\end{align}
\end{widetext}

\subsubsection{Eigenmodes of the C-Slice}\label{subsubsec:eigmodes}

For non-trivial solutions of the wave equation, we require that the determinant of (\ref{eq:wavematrix2}) should be equal to zero. From the secular equation we determine the dispersion relations
of the eigenmodes. For a space with the properties of the $C-$slice, they are simple and degenerate:
\[
\omega^{(i)}_{j}=\pm\sqrt{k_{jx}^{2}+k_{jz}^{2}m_{j}},
\]
for each polarisation $i$ and each slice $j$. As in the previous calculation (\ref{subsec-virtualcalc}), we ignore dispersion by setting $m(z,\omega)=m(z)$. The eigenmodes of the electric field (prior to normalisation) are determined to be
\begin{align}
\vec{e}^{(1)}_{j}=\vec{e}^{(2)}_{j}=\left(-\frac{k_{jz}m_{j}^{2}}{k_{jx}},0,1\right),\nonumber
\\
\quad\vec{e}^{(3)}_{j}=\vec{e}^{(4)}_{j}=\left(0,1,0\right).
\end{align}
The eigenmodes of the magnetic field can be derived directly from the eigenmodes of the electric field via
\[
\vec{h}^{(i)}_{j}=\frac{1}{\omega_{j}}\left(\vec{\mu}_{j}\right)^{-1}\left(\vec{k}_{j}\times\vec{e}^{(i)}_{j}\right),
\]
from which we obtain
\begin{align}
 \vec{h}^{(1)}_{j}  = -\vec{h}^{(2)}_{j} &=&
& \left(0,-\frac{m_{j}}{\omega_{j}}\left(k_{jx}+\frac{k_{jz}^{2}m_{j}^{2}}{k_{jx}}\right),0\right), &
\nonumber \\
 \vec{h}^{(3)}_{j}  = -\vec{h}^{(4)}_{j} &=&
& \left(-\frac{k_{jz}m_{j}}{\omega_{j}},0,\frac{k_{jx}}{\omega_{j}m_{j}}\right).&
\end{align}

\subsubsection{Reflection coefficients in a C-Slice}\label{subsubsec:refleccoeffs}

For an interface consisting of two $C-$slice half-spaces, $\alpha$ and $\beta$, we  can relate the field inside the first layer, and to the immediate left of the interface, with the field inside the second layer, and immediately right of the interface, by the zero-phase transfer matrix $\vec{M} = \vec{M}(\beta)$:
\begin{align}
\left(\begin{array}{c}
E_{s}^{t}\\
0\\
E_{p}^{t}\\
0
\end{array}\right)=\vec{M}\left(\begin{array}{c}
E_{s}^{i}\\
E_{s}^{r}\\
E_{p}^{i}\\
E_{p}^{r}
\end{array}\right)
\end{align}
$E_{s}^{i}$, $E_{s}^{r}$, in this case, represent the $C$-modes incident upon and reflected by the boundary, and $E_{p}^{i}$, $E_{p}^{r}$, the incident and reflected $p$-modes. $E_{s}^{t}$ and $E_{p}^{t}$ are the transmitted $s$ and $p$ modes. The reflection coefficients are then:
\begin{align}
 r_{ss} = \left.\frac{E_{s}^{r}}{E_{s}^{i}}\right|_{E_{p}^{i}=0} =  &\frac{M_{24}M_{41}-M_{21}M_{44}}{M_{22}M_{44}-M_{24}M_{42}}, & \nonumber \\
 r_{sp}  = \left.\frac{E_{s}^{r}}{E_{s}^{i}}\right|_{E_{p}^{i}=0} =   &\frac{M_{21}M_{42}-M_{22}M_{41}}{M_{22}M_{44}-M_{24}M_{42}}, & \nonumber \\
 r_{pp} = \left.\frac{E_{p}^{r}}{E_{p}^{i}}\right|_{E_{s}^{i}=0} =    &\frac{M_{22}M_{43}-M_{23}M_{42}}{M_{24}M_{42}-M_{22}M_{44}}, & \nonumber \\
 r_{ps}  = \left.\frac{E_{s}^{r}}{E_{p}^{i}}\right|_{E_{s}^{i}=0} =   &\frac{M_{24}M_{43}-M_{23}M_{44}}{M_{24}M_{42}-M_{22}M_{44}}.& \label{eq:reflec-coeffs}
\end{align}
Thus the reflection coefficients for this inteface are determined directly from $\vec{M} = \vec{D}(\beta)^{-1}\vec{D}(\alpha)$:
\begin{widetext}
\begin{align}
\vec{M}=\frac{1}{2}\left(\begin{array}{cccc}
\frac{m_{\alpha}}{m_{\beta}^{2}}\left(\frac{k_{\alpha z}}{k_{\beta z}}m_{\alpha}+\frac{\omega_{\alpha}}{\omega_{\beta}}m_{\beta}\right) & \frac{m_{\alpha}}{m_{\beta}^{2}}\left(\frac{k_{\alpha z}}{k_{\beta z}}m_{\alpha}-\frac{\omega_{\alpha}}{\omega_{\beta}}m_{\beta}\right) & 0 & 0\\
\frac{m_{\alpha}}{m_{\beta}^{2}}\left(\frac{k_{\alpha z}}{k_{\beta z}}m_{\alpha}-\frac{\omega_{\alpha}}{\omega_{\beta}}m_{\beta}\right) & \frac{m_{\alpha}}{m_{\beta}^{2}}\left(\frac{k_{\alpha z}}{k_{\beta z}}m_{\alpha}+\frac{\omega_{\alpha}}{\omega_{\beta}}m_{\beta}\right) & 0 & 0\\
0 & 0 & 1+\frac{k_{\alpha z}\omega_{\beta}}{k_{\beta z}\omega_{\alpha}}\frac{m_{\alpha}}{m_{\beta}} & 1-\frac{k_{\alpha z}\omega_{\beta}}{k_{2z}\omega_{\alpha}}\frac{m_{\alpha}}{m_{\beta}}\\
0 & 0 & 1-\frac{k_{\alpha z}\omega_{\beta}}{k_{\beta z}\omega_{\alpha}}\frac{m_{\alpha}}{m_{\beta}} & 1+\frac{k_{\alpha z}\omega_{\beta}}{k_{2z}\omega_{\alpha}}\frac{m_{\alpha}}{m_{\beta}}
\end{array}\right).
\end{align}
\end{widetext}
Further simplification of this matrix is possible, since $\omega_{\alpha}=\omega_{\beta}$ and
\begin{align*}
\sqrt{k_{\alpha x}^{2}+k_{\alpha z}^{2}m_{\alpha}^{2}}=\sqrt{k_{\alpha x}^{2}+k_{\beta z}^{2}m_{\beta}^{2}} \\
\implies k_{\alpha z}^{2}m_{1}^{2}=k_{\beta z}^{2}m_{\beta}^{2}.
\end{align*}
We know that these factors share the same signs. Therefore we infer that
\[
k_{\alpha z}m_{\alpha}=k_{\beta z}m_{\beta}.
\]
This, of course, is to be expected: the $k_{jz}$ wave numbers in a homogeneous slice $j$ are equal to the vacuum wave number $k_0z$ compressed by a factor of $m_{j}$,
\[
 k_{jz} = \frac{k_{0z}}{m_{j}}.
\]
With these simplications, it is clear that the matrix is diagonal, and reduces to the form
\begin{align}
\vec{M}=\left(\begin{array}{cccc}
\frac{m_{\alpha}}{m_{\beta}} & 0 & 0 & 0\\
0 & \frac{m_{\alpha}}{m_{\beta}} & 0 & 0\\
0 & 0 & 1 & 0\\
0 & 0 & 0 & 1
\end{array}\right).
\end{align}
Thus we find that, for all angles of incidence, the reflection coefficients (\ref{eq:reflec-coeffs}) are identically zero; there is no internal scattering within a C-Slice.

\subsubsection{The force on the plate}

	\par
	If we consider now the whole system, consisting of a series of $N+1$ homogeneous slices between and contiguous with the two mirrors, and the reflection coefficients associated with sending plane waves to the right and to the left of any given point $x_{l}$ in the cavity,  it is clear that they are of the same form (\ref{eq:reflec-coeffs}), but with the matrix $\vec{M}$ substituted by the transfer matrix $\vec{T_{L}}$ or $\vec{T_{R}}$, when calculating the left or right reflection coefficients respectively. Between the plates, these transfer matrices are diagonal matrices consisting of phase terms, being made up of products of diagonal $\vec{M}$ and $\vec{\Phi}$ matrices. For $j \in [2,N]$, the $\vec{M}$ matrices do not modify the reflection coefficients, and can be replaced with identity matrices, hence the contribution to the transfer matrices for $j \in (2,N)$ consists entirely of phase terms that divide into two sets: the phase terms for which $m_{j}=1$ (ie. slices of vacuum), and the phase terms for which $m_{j} \ne 1$ (ie. $C-$slice material). The reflection matrices (\ref{eq:reflecmatrices}) are diagonalised.
	\par
	We recover Casimir's original result by setting $m_{j}=1 \, \forall j$. This problem has already been solved, and we will not repeat the details of that calculation again. To determine the force on the left or right plate in our example, where a $C-$slice has been introduced into the cavity, we need only consider the way in which the $C-$slice modifies the relevant transfer matrices that determine the reflection coefficients. From the discussion above, it is evident without further calculation that this difference amounts to nothing more than a modification of the accumulated (imaginary) phase: slices containing $C-$slice material produce an amount of phase in their corresponding transfer matrices that differs by a factor of $m_j$ from slices of vacuum, modifying the $e^{2d w}$ term in the stress to $e^{2d' w}$~\footnote{The $m$ factors that appear in the $M$ matrix disappear in any product of the transfer matrices spanning two sides of the device, which is situated in vacuum, where $m=1$}. The reflection coefficients at the mirrors (\ref{eq:plate-coeffs}) remain unmodified during the motion. This result is consistent with our prediction using a simpler argument from transformation optics, producing precisely the same expression for the Casimir force (\ref{eq:newCasimir}). The paradox has happily been averted.

\section{Applications of the $C-$slice}

	\par
	Quantum stiction, due to the `stickiness' of the Casimir effect, has been acknowledged as a serious engineering problem for micro and nano--machinery~\cite{capasso2009}. At such length scales, Casimir-Lifshitz forces are no longer negligible and can lead to unwanted attraction between, and adhesion of, material parts in the device. Although immersion in a suitably dense liquid may reduce the phenomenon of quantum stiction (by effectively increasing the optical distance between the parts) this would introduce viscous forces into the system, among other complications. For micromachines that cannot be thus immersed (for example, where the parts need to move in relation to each other) the freedom of a surrounding vacuum remains necessary. An idealised $C-$slice seems to suggest itself as one possible solution to this problem. Without modifying the physical dimensions of the cavity, or introducing new surfaces for interaction, the {\em effective} size of the cavity can be made arbitrarily large, and the Casimir force made arbitrarily small, by interposing a thin wafer made to the appropriate specifications (\ref{eq:specs})\footnote{In this case, a homogeneous $C-$slice would be sufficient to achieve the desired effect, and should be engineered so that $m < 1$}. A similar proposal for tackling stiction was made in \cite{leonhardt2007}, using a negatively refracting medium to produce a repulsive Casimir force, with the disadvantage that optical pumping would be required. The $C-$slice does not require optical pumping.  However, as we observed earlier, the quantum Casimir effect is a broadband phenomenon, and it seems unlikely that the necessary condition of impedance-matching could be secured for a sufficiently large portion of the electromagnetic spectrum, though an argument to the contrary can be found in \cite{leonhardt2007}. Whether or not a material could be made that approximated the effects of the $C-$slice is a problem we will not examine here. For the thermal Casimir effect, however, it has been observed that, as the temperature increases, contributions to the force are increasingly distributed around a characteristic frequency~\cite{miao2010}; it is perhaps more conceivable that an impedance-matched metamaterial implementing the $C-$slice could be designed for tuning the thermal Casimir effect.

	\par
	However, there are applications outside of the domain of Casimir Physics.  By virtue of its non-reflective properties and its capacity to change the measure of optical distance, the compressive transformation implemented by a $C-$slice can be used to reduce the profile of a lens \cite{Roberts2009}, for example. Perhaps there may also be some use for the $C-$slice in a laser system, should there be occasion to modify the resonant frequency of a cavity without disturbing the cavity walls. Further development of these ideas is beyond the scope of this paper, however.

\section{Conclusions}

        \par
	We have imagined introducing an inhomogeneous transformation medium into a cavity that implements a simple distortion of the laboratory coordinate system. Initially, it appeared that there may be a contradiction between the force predicted by transformation optics in the virtual space of the system, which is finite and cutoff-independent, and the force predicted by Lifshitz theory in the physical space of the laboratory, since Casimir forces in inhomogeneous media have generally been found to be cutoff-dependent~\cite{simpson2013,horsleysimpson2013}.
	\par
	However, we have established that there is no contradiction here. The Casimir force in an inhomogeneous medium typically depends on microphysical details which, as yet, remain unincorporated within Lifshitz theory; in such cases, where there is scattering within the medium, the integrand of the stress diverges~\cite{simpson2013} and the force is cutoff-dependent~\cite{horsleysimpson2013}. However, in the case of optical devices that implement virtual geometries for light, the reflection coefficients are precisely zero for all angles of incidence, and the stress does not diverge. A tranformation medium merely changes the measure of optical distance. Therefore, it is possible to determine the Casimir forces for systems incorporating transformation media, even when they involve continuously changing optical properties, and we have identified a subset of inhomogeneous systems for which Lifshitz theory may still make meaningful predictions \footnote{The only other example where a finite Casimir force has been calculated for an inhomogeneous medium that the author is aware of is \cite{leonhardt2011}. In this case, however, the inhomogeneous system did not constitute a transformation medium, and Lifshitz' regularisation was not applied.}. In the example we have considered, where we interpose a $C-$slice between two parallel plates, the $C-$slice changes the Casimir force by modifying the effective distance of the cavity, but the force between the plates remains attractive and finite.

%
%
	\acknowledgements
	WMRS wishes to thank Jiaming Hao for a helpful discussion, Simon Horsley for suggesting the basic problem and critically examining a draft of the paper, and Ulf Leonhardt, who also offered helpful feedback on this paper. The financial support of SUPA and the Weizmann Institute is gratefully acknowledged.

\end{document}